\documentstyle[epsf,psfig,prl,multicol,aps]{revtex}
\begin{document}
\title{Prediction of High T$_c$ Superconductivity in Hole-doped LiBC} 
\author{H. Rosner, A. Kitaigorodsky, and
W. E. Pickett}
\address{
Department of Physics, University of California, Davis CA 95616 \\
}

\date{\today}
\maketitle
\begin{abstract}
The layered lithium borocarbide LiBC, isovalent with and structurally
similar to the superconductor MgB$_2$, is an insulator due to the
modulation within the hexagonal layers (BC {\it vs.} B$_2$).
We show that hole-doping of LiBC results in Fermi surfaces of B-C
$p\sigma$ character that couple very strongly to B-C bond stretching
modes, precisely the features that lead to 
superconductivity at T$_c \simeq$ 40 K in MgB$_2$.
Comparison of Li$_{0.5}$BC with MgB$_2$ indicates the former to
be a prime candidate for electron-phonon coupled
superconductivity at substantially higher temperature than in MgB$_2$.
\end{abstract}

\begin{multicols}{2}
The discovery of $\sim$40 K superconductivity in
MgB$_2$~\cite{akimitsu} has led to a resurgence of interest in layered
compounds that are metallic, or may be made metallic by doping.
MgB$_2$ itself is, to date, a unique system, one that a number of
studies of the electronic
structure\cite{tupitsyn,jan,kortus,ajf,satta,bela,nishibori,psingh}
confirm contains strongly covalent bonds similar to graphite yet is a
good metal with strong electron-phonon (EP) coupling.  A variety of
evaluations of the EP coupling indicates that it is
strong enough to account for superconductivity in the 40 K
range,\cite{jan,kortus,kong,bohnen,yildirim,liu} and the observed B
isotope shift of T$_c$\cite{budko} confirms that pairing is due to
phonons.  So far, attempts to make related, nearly isoelectronic
diborides with the same structure have been fruitless.  Be
``diboride'' (actually BeB$_{2.75}$) forms a complex compound\cite{young}
distinct from the MgB$_2$ type structure that has been presumed for
several band structure studies.\cite{ajf,satta,psingh} Ca, Sr, and Ba
have resisted incorporation into the MgB$_2$ structure.  Al can be
alloyed in, resulting in a decrease in T$_c$ and a subsequent collapse
of superconductivity altogether.  Alloying with monovalent ions for Mg
should increase T$_c$ (based on rigid band concepts) but attempts to
do this also have been unsuccessful.

A closely related compound that has not received attention since the
discovery of MgB$_2$ is LiBC.  Although apparently 
straightforward to make,\cite{worle} it has been 
studied very little to date.  Its
structure\cite{worle,mair,ramirez} can be derived from the ``fully
intercalated graphite'' structure of MgB$_2$ by Mg $\rightarrow$ Li,
and B$_2 \rightarrow$ BC, with the hexagonal BC layers alternating so
that B is nearest neighbor to C along the $\hat c$ axis as well as
within the layer.  Like the compound CuBC suggested by Mehl {\it et al.}
\cite{mehl}, it is isovalent with MgB$_2$: Li has one less
valence electron than Mg, but C has one more than B.  The main
features of the electronic structures, which are dominated by in-plane
B-B and B-C bonding respectively, should be 
expected to have strong similarities.
LiBC is insulating, however, with a gap that (we show below) arises
from the replacement B$_2$ $\rightarrow$ BC, and not from the 
alternating stacking.  An important feature of
Li$_x$BC is that the Li content can be varied within the same
crystal structure:  
while this system forms stoichiometrically ($x$=1),
the Li content can be reduced and a
structural change was seen only for $x$=0.24.\cite{worle}

Ramirez {\it et al.}\cite{ramirez} have presented semi-empirical
Hartree-Fock (INDO) results for a (BC)$^{-}$ layer representing LiBC
in the absence of interlayer coupling.  They obtained a band gap of
4.3 eV (acknowledged to be an overestimate due
to neglect of correlation), with its minimum value at
the K point of the Brillouin zone between bonding and antibonding
$\pi$ ($p_z$) bands.  B and C character of the bands were not distinguished
in their results.  The $\sigma$ bands ($s p_x p_y$ hybrids)
however lie above the $p\pi$
valence bands in a region around the $\Gamma$ point.  Albert included
LiBC in a discussion\cite{albert} of Li borocarbides, most of which
are structurally very different from LiBC.

In this paper we present results for the electronic structure 
and EP coupling of Li$_x$BC ($x$
= 0.5, 0.75, 1.0) and compare
it with that of MgB$_2$.  We determine that once it is hole-doped, the
states at the Fermi level include $\sigma$ bands
with cylindrical $\sigma$ Fermi surfaces that are strongly coupled to 
the B-C bond-stretching mode (more strongly than in MgB$_2$).  
At $x \simeq$ 0.5, the density of states (DOS)
N(E$_F$) is comparable to that of MgB$_2$ while the EP coupling is
much stronger.  These features suggest that
hole-doped LiBC is a very favorable candidate for
superconductivity, possibly at much higher temperature than in
MgB$_2$.

\begin{figure}[bt]
\begin{center}

\psfig{figure=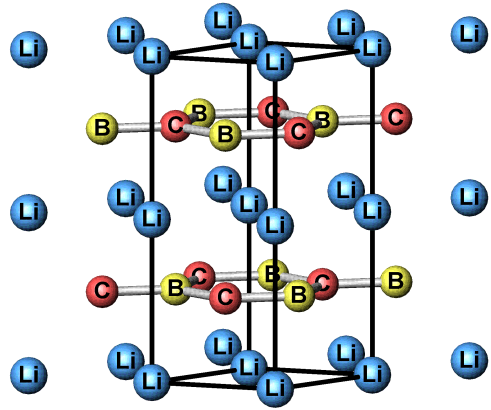,width=6.5cm}
\end{center}
\vspace{3mm}
\caption{
Crystal structure of LiBC, a direct generalization of that of MgB$_2$,
showing the alternating layers of
hexagonal BC. Li ions lie in the interstitices.
}
\label{libc}
\end{figure}

The layered structure of LiBC, pictured in Fig. 1, has been determined
by W\"orle {\it et al.}\cite{worle} from golden hexagonal platelets
(and red in transmission, hence insulating) to be P6$_3/mmc$
(D$_{6h}^4$, No. 194), $a$= 2.752 \AA, $c$ = 7.058 \AA = 2$\times$
3.529 \AA.  This
structure is similar to that of MgB$_2$, except that $a$ is 9\% smaller;
the interlayer spacing is nearly identical.
The BC layers are stacked in alternating fashion, so
there is no B-B or C-C bonding in the structure.
To expose the effects of A-B versus A-A stacking of the layers, we also
consider the A-A stacking, which has space group P$\bar 6m2$ (D$_{3h}^1$,
No. 187).  

\end{multicols}
\begin{figure}[tbp]
\epsfxsize=17cm\centerline{\epsffile{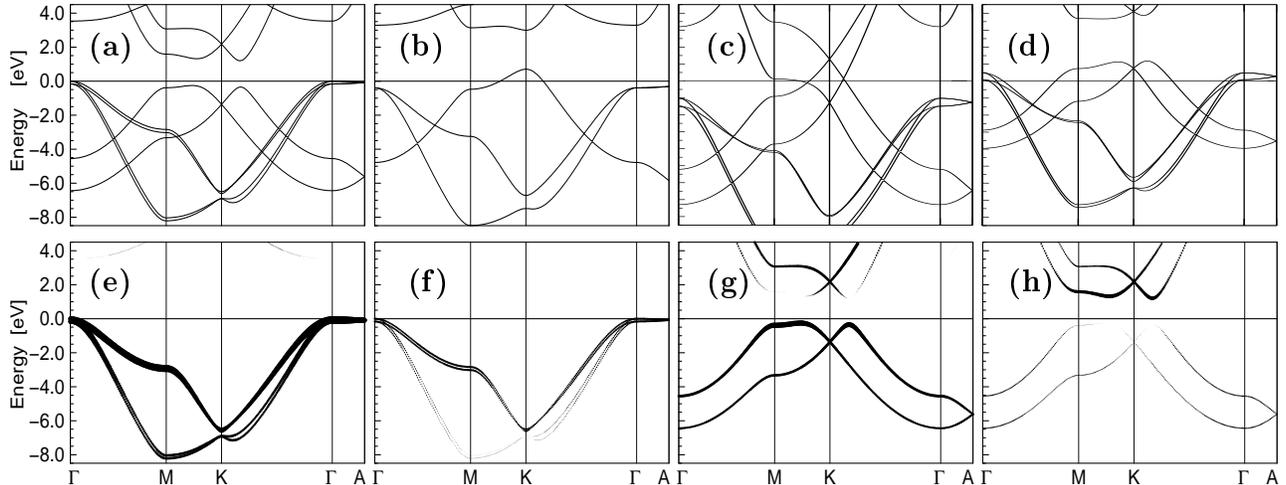}}
\caption{Band structures of:
(a) LiBC in the observed structure;
(b) LiBC in fictitious A-A stacking structure, where additional $k_z$
   dispersion produces a semimetallic band structure;
(c) Li${\cal B}_2$, the virtual crystal analog;
(d) Li$_{0.5}$BC in virtual crystal approximation;
(e) C $p\sigma$ ``fatbands;
(f) B $p\sigma$ ``fatbands;
(g) C $p\pi$ ``fatbands;
(h) B $p\pi$ ``fatbands.
}
\label{fatbands}
\end{figure}
\begin{multicols}{2}

We have used two electronic structure methods.
A full-potential nonorthogonal local-orbital scheme
\cite{koepernik99} was used to
obtain accurate band structures and the total energies that we report.
Li(2$s$, 2$p$, 3$d$), B (2$s$, 2$p$, 3$d$) and C (2$s$, 2$p$, 3$d$)
states, respectively, were chosen as the basis set. All lower lying
states were treated as core states.  The Li 2$p$ and 3$d$ states as
well as the B and C 3$d$ states were taken into account to increase
the completeness of the basis set. The spatial extension of the basis
orbitals, controlled by a confining potential \cite{eschrig89}
$(r/r_0)^4$, was optimized to minimize the total energy.  In addition,
the LMTO-47 code\cite{LMTO} was used for the virtual crystal calculations
and to obtain the deformation potentials that we report below.
Atomic sphere radii of 3.00 a.u. (Li) and 1.73 a.u.
(B and C) were used.  


{\it Stoichiometric Compound.}
The band structure of LiBC (observed structure) along high 
symmetry lines in the Brillouin
zone (BZ) is shown in Fig. 2(a).  Valence bands are separated from
conduction bands by a calculated gap of 1.0 eV, with the valence band
maximum occurring at $\Gamma$, and the conduction band minimum
occurring at H.  As in other semiconductors, this LDA value is
likely to be an underestimate of the experimental gap.  It will be
shown below that the band structure of LiBC is closely related to that
of MgB$_2$, so the first question to address is the origin of the band
gap at the Fermi level in LiBC, with the two possibilities being the
alternating stacking of B-C layers along the $\hat c$ axis, and the
alternation of B and C within the layer itself.

This question is answered by performing two calculations, one for a
isoelectronic virtual crystal ``Li${\cal B}_2$,'' where ${\cal B}$ is
an average atom between B and C with nuclear charge Z=5.5, and another
for an alternative LiBC structure in which the stacking of layers
along the $\hat c$ axis is A-A stacking rather than the observed A-B
stacking. (This is the structure for CuBC considered by Mehl {\it et al.}
\cite{mehl}). These results are also shown in Fig. 2(b)-(c).  The bands of
Li${\cal B}_2$ are isomorphic to those of MgB$_2$, as expected, as it
is structurally identical and chemically similar.  The bands look
different from those normally shown for MgB$_2$ because we have used a
doubled cell (along $\hat c$) to allow direct comparison with the
bands of LiBC.  The $\sigma$ and $\pi$ bands (see below) appear twice
due to BZ folding, with the former separated by only a few tenths of
eV due to their layered character ($s-p_x-p_y$) and the latter ($p_z$)
separated by 3-4 eV around and below the Fermi level E$_F$.

The A-A stacking of BC layers [Fig.~2(b)] produces a 
substantial change of the
electronic structure: it does not change the separation of the valence
and conduction bands, but it does change $k_z$ dispersion sufficiently that
the gap disappears -- it becomes a semimetal with a calculated band
overlap of 1.7 eV.  
The calculated total energy of the A-B stacking is lower than A-A 
stacking (at constant
structure, the experimental one) by 35 meV per formula unit, 
consistent with the observed A-B stacking.

To understand the electronic structure of LiBC and its similarity to
that of MgB$_2$ more clearly, we display in Fig. 2(e)-(h) the ``fatbands''
plots showing C and B $\sigma$ and $\pi$ character. 
The $\sigma$ bands are as in MgB$_2$, with C character concentrated
more heavily at the bottom of the bands (-6 to -8 eV below the gap)
because the on-site energies $\varepsilon_s$ and $\varepsilon_p$ 
are lower for C than for B.  
The gap in LiBC, also evident in the DOS plots in Fig. 3,
lies between bonding and antibonding combinations of
$\pi$ bands that are forced by symmetry in MgB$_2$ to connect, and
thus enforce metallicity.  The bonding $\pi$ bands have a
preponderance of C $p_z$ character, while the antibonding bands just
above the gap are more strongly B $p_z$ in character.  
The ionic character of the B-C layer compared to B-B helps to push 
the $\pi$ character downward with respect to the $\sigma$ character,
resulting in a larger proportion of $\sigma$ character 
[N$_{\sigma}(\varepsilon)$] than in MgB$_2$, a point of some
importance below.

\relax
\begin{figure}[tbp]
\begin{center}
\begin{minipage}{5.6cm}
\psfig{figure=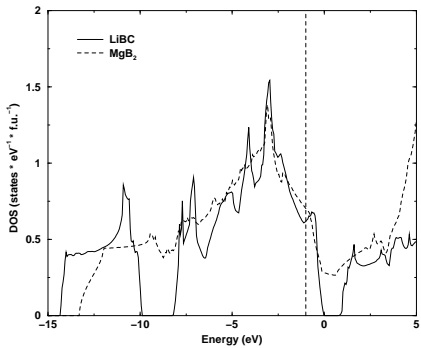,angle=-90,width=5.5cm}
\end{minipage}
\end{center}
\caption{The density of states of LiBC (solid line) compared to that of
MgB$_2$ (dashed line, with the vertical line denoting the Fermi level);
the gaps arise from in-plane (B-C) modulation.
The main valence band peaks have been aligned to illustrate the
similarity.
}
\label{dos1}
\end{figure}

{\it Substoichiometric Li$_x$BC.}
In a band semiconductor such as LiBC, heavy hole doping 
will make Li$_x$BC metallic.  If the
$\sigma$ bands can be populated by a sufficient density of holes,
then by analogy with MgB$_2$ the system should become superconducting,
and perhaps a very good superconductor.  There are at least two
crucial questions.  First, can the Fermi level be moved far enough into
the valence bands to give a value of N(E$_F$)
large enough to be interesting --
comparable to that of MgB$_2$, say?  Second, are these $\sigma$
states strongly coupled to the B-C bond stretching modes as 
they are in MgB$_2$?

We have used supercells (doubled within the layers) to study the $x$ = 0.5
and $x$ = 0.75 concentrations.  For the $x$ = 0.5 case, two
configurations were studied to assess the effect of the placement of
missing Li atoms.  In both configurations half of the 
Li atoms were removed from each
layer, in one case along a line parallel to the $\hat c$ axis, and
in the other case in staggered positions.  The differences are very minor.
In Fig. 4 the resulting DOS of these systems are compared with those
for $x$ = 1.  The Fermi level moves into the valence band, as it must,
and the behavior in the region of the gap and Fermi level is not far
from rigid band.  It is noteworthy that it is the $\sigma$ bands that are
occupied first as holes are doped in [Fig. 2(a)].
The value of N(E$_F$) = 0.68 eV$^{-1}$ at $x$ = 0.5
is nearly equal to that of MgB$_2$. 

\begin{figure}[tbp]
\epsfxsize=6.0cm\centerline{\epsffile{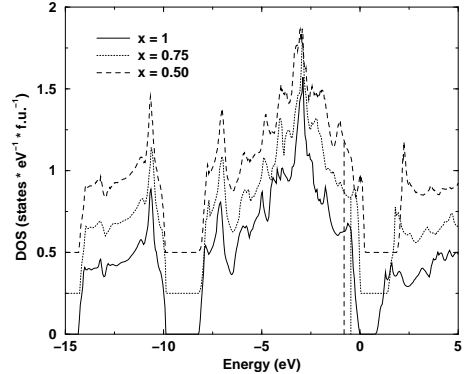}}
\caption{Densities of states of Li$_x$BC, for $x$=1, 0.75, and 0.5,
displaced upward consecutively for clarity.  The $x$=0.5 curve is
averaged over two possible supercells (see text).
The main valence band peaks have been aligned (but the gaps also
remain aligned), so the Fermi level (vertical lines) moves
downward, roughly in rigid band fashion, as hole doping proceeds.
}
\label{dos2}
\end{figure}

{\it Estimation of T$_c$.}
EP coupling (strength $\lambda$) in MgB$_2$, 
which drives its superconductivity,
is dominated by coupling of B-B bond stretching modes to the $\sigma$
cylinder sheets of Fermi surface.\cite{jan,kortus,kong,bohnen,yildirim,liu}
We calculate the deformation potential of these
states for the zone-center B-C bond stretching mode in Li$_{0.5}$BC
to be ${\cal D}$ = 18.5 eV/\AA, more than 40\% larger than in MgB$_2$
(where it is 13 eV/\AA \cite{jan}), resulting from the stronger B-C
bonding compared to B-B.  We calculate $\omega_{E_{2g}}$ = 68 meV for
Li$_{0.5}$BC.  We can obtain a realistic estimate
using\cite{jan} $\lambda \propto$ 
N$_{\sigma}$(E$_F$) ${\cal D}^2/M\omega^2$ and scaling from calculated
quantities and experimental T$_c$ for MgB$_2$.
Defining the ratio ${\cal R}(K)
= K_{Li_{0.5}BC}/K_{MgB_2}$ of any property $K$, 
our calculations give ${\cal R}
(N_{\sigma})=1.3$, ${\cal R}({\cal D}^2) = 2.0,$ and
${\cal R}(\omega_{E_{2g}}) = 1.17,$ finally giving ${\cal R}(\lambda) = 2.1$
as the enhancement of $\lambda$ in Li$_{0.5}$BC compared to that of MgB$_2$.  

The value of $\lambda$ in MgB$_2$, averaged over four 
groups who have done extensive calculations,\cite{kong,bohnen,yildirim,liu}
is 0.82$\pm$0.08.  The enhancement of 2.1 (above) gives 
$\lambda \simeq 1.75$ for Li$_{0.5}$BC.  We obtain T$_c$ from the Allen-Dynes
equation\cite{pba} with a single frequency; choosing $\mu^* = 0.09$ with our
calculated $\omega$ = 58 meV (similar to \cite{kong,bohnen,yildirim,liu})
and $\lambda = 0.82$ gives T$_c$ = 39 K for the reference material MgB$_2$,
as observed. 
Using the same $\mu^*$, our calculated value $\omega_{E_{2g}} 
= 68$ meV, and $\lambda$ = 1.75
gives T$_c \simeq$ 115 K for $x$=0.5.  Variation with $x$ should be 
dominated by the $\sigma$ DOS N$_{\sigma}$(E$_F$($x)$), and in Fig. 5 we
plot the calculated T$_c$($x$) assuming other quantities are constant
to give an idea of the rapidity of the variation of T$_c$($x$).
This estimate indicates that T$_c$ should be higher than 40 K for
$x \geq$ 0.25.  This estimate neglects multiband effects which are
substantial for MgB$_2$,\cite{multi} but our purpose of illustrating the
likely value of T$_c$ relative to MgB$_2$ should be reasonable.
Even if our method of estimating T$_c$
overoptimistic, Li$_x$BC should still
be an impressive superconductor.

\begin{figure}[tbp]
\psfig{figure=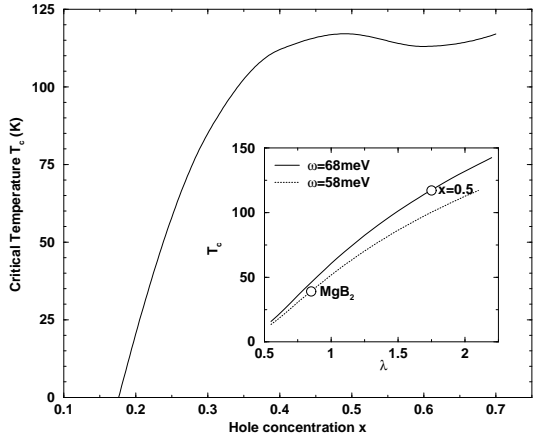,angle=-90,width=7.0cm}
\caption{Calculated T$_c$ versus hole concentration, but accounting
only for the $x$ dependence of N$_{\sigma}$(E$_F$).  The  curve
is based on the
model described in the text that is pegged to reproduce MgB$_2$
exactly with $\lambda$=0.82, $\mu^*$ = 0.09, $\omega$ = 58 meV.
The inset shows the $\lambda$ dependence of T$_c$ for MgB$_2$
(calculated $\omega$ = 58 meV) and Li$_{0.5}$BC
(calculated $\omega$ = 68 meV).
}
\label{Tc}
\end{figure}

According to W\"orle {\it et al.}\cite{worle} the hole doping we have 
treated has already been achieved, but measurements at reduced
temperature were not made.  In the event that doping is difficult,
LiBC (being a semiconductor) seems to be a prime 
candidate for field-effect doping (FED) to achieve
superconductivity.  FED has already been reported
to produce (1) T$_c$ = 52 K
in field-effect hole-doped C$_{60}$ and at T$_c$ = 117 K 
in expanded C$_{60}$\cite{batlogg1}, and (2) at 14 K in a ladder
cuprate and 89 K in an ``infinite layer'' 
cuprate\cite{batlogg2}.  Not only do 
these studies establish the capabilities of the FED technique, 
they demonstrate that phonon-coupled superconductivity (as in the
fullerenes) can produce
very high values of T$_c$, as we are suggesting here for Li$_x$BC.

EP coupling of the strength that we obtain for Li$_x$BC
must raise the question of the associated lattice instabilty that will
ultimately occur.  In MgB$_2$ the $E_{2g}$ mode remains stable although
it acquires a very large width\cite{bohnen} and no doubt is renormalized
downward strongly\cite{allen} by the large coupling 
to the $\sigma$ states.  [In
weakly coupled AlB$_2$ $\omega_{e_{2g}}$ = 123 meV.\cite{bohnen}]  With its
much stronger coupling, Li$_x$BC provides the opportunity to approach --
and perhaps encounter -- the associated instabilty.  We are currently 
calculating $\omega_{E_{2g}}(x)$ to study this instability; note however
that supposing a (say) 20\% lower frequency actually leads to a higher
T$_c$ in this regime, because the larger value of 
$\lambda$ more than compensates the
decrease in energy/temperature scale.
Raman and inelastic
neutron spectroscopy will be very informative in revealing the response
of the lattice to this large EP coupling.

To gather these results together: we have shown that in comparison to
MgB$_2$, the material Li$_x$BC, $x \sim$ 0.5, has (i) the hole-doped
$\sigma$ states at E$_F$ that are crucial in MgB$_2$, (ii) N(E$_F$)
is comparable to that of MgB$_2$, (iii)
an E$_{2g}$ deformation potential of
the $\sigma$ bands that is over 40\% larger than in MgB$_2$, and (iv) the
E$_{2g}$ mode frequency is similar to that in MgB$_2$.  In combination,
these results strongly suggest a coupling strength perhaps as high as
$\lambda \simeq$ 1.75,
and a superconducting critical temperature
that could be more than twice as high as in MgB$_2$ for $x = 0.5 \pm 0.2$..

We thank J. M. An for calculation of the phonon frequencyof Li$_{0.5}$BC.
This work was supported by National Science Foundation Grant
DMR-9802076, and by the Deutscher Akademischer Austauschdienst.

\vskip -7mm

\end{multicols}
\end{document}